\documentstyle[aps,preprint]{revtex}

\begin{document}
\draft
\title{ Fermi liquid behavior and spin-charge separation in underdoped cuprates}
\author{Ching-Kit Chan and Tai-Kai Ng}
\address{
Department of Physics,
Hong Kong University of Science and Technology,
Clear Water Bay Road,
Kowloon, Hong Kong
}
\date{ \today }
\maketitle
\begin{abstract}
  The Gaussian fluctuations in slave-boson mean-field theory of $t-J$ model is analyzed in this paper in the
  low-doping regime where the superconducting to normal (pseudo-gap phase) transition is driven by vanishing of
  bose-condensation amplitude. By eliminating the boson and constraint fields exactly in the linear response regime
  we show that the Gaussian theory describes a Fermi liquid superconductor where the superconducting to normal
  transition is actually a spin-charge separation transition characterized by a change of Landau parameter
  from $F_1>-1$ in the superconducting phase to $F_1=-1$ in the pseudo-gap phase. Consequences of this proposal are
  discussed.

\end{abstract}

 \pacs{PACS Numbers: 71.10.Fd, 71.10.Hf, 71.27.+a, 74.20.Mn, 74.72.-h}

\narrowtext

   With accumulated experimental evidences it is now generally accepted that at low temperature high-$T_c$
  superconductors are BCS-like d-wave superconductors close to Mott insulator\cite{e1,e2,e3}. It is believed that
  strong electron correlation does not destroy Fermi-liquid behaviour, and the ground state of the superconducting
  states remains Fermi-liquid-(superconductor)\cite{e3,t1} like. However the situation is much less clear in the
  normal state (pseudo-gap regime) of underdoped cuprates where a d-wave BCS-like gap seems to survive, although the
  system is already a normal metal. Photoemission experiment indicates that a segmented Fermi surface seems to exist
  in this regime\cite{e1,p1} which is hard to reconcile with a Fermi liquid description. A theoretical challenge is
  thus whether a phenomenological Fermi-liquid type description can be found for the pseudo-gap phase and what is
  the nature of the corresponding superconducting to normal transition?

     The purpose of our paper is to address this question based on the $U(1)$ slave-boson mean field theory (SBMFT) of
  $t-J$ model\cite{sbtj,t2,t3,greview}. SBMFT provides an unconventional description of the high-$T_c$ cuprates
  where the pseudo-gap phase is described as a state with a d-wave spinon gap but superfluidity vanishes because of
  vanishing (slave)-boson condensation. This description is suggestive but not entirely satisfactory because of the
  explicit appearance of slave-degree of freedom in the theory. The description would be much more convincing if
  the physics can be expressed in terms of physical degrees of freedom, or quasi-particles only, without referring
  to slave particles.

   A previous study\cite{ng1} indicates that at zero temperature the slave-degrees of freedom (constraint field and
  slave bosons) in SBMFT can be eliminated exactly in Gaussian theory, resulting in a linear transport equation for
  quasi-particles which has the same form as transport equation for Fermi-liquid superconductors\cite{leggett}, with
  all Landau interaction parameters explicitly given, indicating that SBMFT describes a conventional Fermi-liquid
  superconductor at zero temperature\cite{ng1}. We shall generalize this approach to finite temperatures with arbitrary
  bose-condensation amplitude in this paper. Our goal is to obtain a formulation of SBMFT in terms of physical
  particles only, and to see how the Landau Fermi-liquid description which comes out naturally in
  the superconducting state, is modified in the pseudo-gap state where bose-condensation amplitude vanishes.

  We consider a generalized $t-J$ model on a square lattice\cite{sbtj} that includes Coulomb interaction between
 charges\cite{tjdlee}. In slave-boson representation the Hamiltonian is
 \begin{eqnarray}
 \label{ham}
 H & = & -t\sum_{<i,j>\sigma}\left(b_ib^+_jc^+_{i\sigma}c_{j\sigma}+H.C.\right)-
 \mu\sum_{i\sigma}c^+_{i\sigma}c_{i\sigma}+J\sum_{<i,j>}\vec{S}_i.\vec{S}_j
 +{1\over2}\sum_{i,j}V(\vec{r}_i-\vec{r}_j)b^+_ib_ib^+_jb_j  \\  \nonumber
 & & +\sum_{i}\lambda_i\left(b^+_ib_i+\sum_{\sigma}c^+_{i\sigma}c_{i\sigma}-1\right),
 \end{eqnarray}
 where $c^+_{i\sigma}(c_{i\sigma})$ and $b^+_i(b_i)$ are the spin (fermion) and hole (boson) creation (annihilation)
 operators at site $i$, respectively. The electron annihilation operator is given by $f_{i\sigma}=c_{i\sigma}b^+_i$.
 $\vec{S}_i=\psi_i^+\vec{\sigma}\psi_i$, where $\vec{\sigma}$ are Pauli matrices and
 $\psi_i=\pmatrix{c_{i\uparrow} \cr c_{i\downarrow}}$. The first term in the Hamiltonian represents electron hopping
 where $<i,j>$ denotes nearest neighbor pairs on a square lattice and $\mu$ is the chemical potential. The third and
 forth terms represent Heisenberg exchange interaction between electron spins and repulsive Coulomb interaction
 between charges, respectively. The non-double occupancy constraint is enforced by the last term
 which introduces a Lagrange multiplier field $\lambda_i$ to the Hamiltonian.

   Following \cite{sbtj} we decouple various terms in the Hamiltonian as follows:
 \begin{mathletters}
 \label{decoupling}
 \begin{equation}
 \label{det}
 b_ib^+_jc^+_{i\sigma}c_{j\sigma}\rightarrow<b_ib^+_j>c^+_{i\sigma}c_{j\sigma}+b_ib^+_j<c^+_{i\sigma}c_{j\sigma}>-
 <b_ib^+_j><c^+_{i\sigma}c_{j\sigma}>,
 \end{equation}
 where $<..>$ denotes expectation value. $<b_ib^+_j>=<b>^2=x$ at zero temperature, where $x=$ concentration of
 holes, but has to be determined self-consistently in general\cite{sbtj}. In particular $<b>=0$ but
 $<b_ib^+_j>\sim x\neq0$ above $T_c$ (pseudo-gap state). Similarly,
 \begin{eqnarray}
 \label{dej}
 \vec{S}_i.\vec{S}_j & \rightarrow & -{3\over8}\left[<\Delta^+_{ij}>\Delta_{ij}
 +\Delta^+_{ij}<\Delta_{ij}>-<\Delta^+_{ij}><\Delta_{ij}>\right.  \\  \nonumber
 & & +\left.<\chi^+_{ij}>\chi_{ij}+\chi^+_{ij}<\chi_{ij}>-<\chi^+_{ij}><\chi_{ij}>\right],
 \end{eqnarray}
 where $\Delta_{ij}=c_{i\uparrow}c_{j\downarrow}-c_{i\downarrow}c_{j\uparrow}$
 and $\chi_{ij}=\sum_{\sigma}c^+_{i\sigma}c_{j\sigma}$. The boson (density-density) interaction term is decoupled as
 $n_in_j\rightarrow<n_i>n_j+n_i<n_j>-<n_i><n_j>$, where $n_i=b^+_ib_i$\cite{tjdlee} and the Lagrange multiplier field
 is replaced by a number function, $\lambda_i\rightarrow<\lambda_i>$ in mean-field theory.
 \end{mathletters}

 A time-dependent SBMFT can be formulated by considering the Heisenberg equation of motion
 ${\partial\over\partial{t}}\hat{O}=i[H,\hat{O}]$ $(\hbar=1)$ for operators $\hat{O}_2$'s that are quadratic in
 the fermion creation/annihilation operators. The equations of motion for $<\hat{O}_2>$ will generate
 terms $<\hat{O}_4>$ which involve four-operator terms. The expectation values of the four-operator terms are
 decoupled in terms of products of two-operator terms in the equation of motion, i.e. $<\hat{O}_4>\sim
 <\hat{O}_2^{(1)}><\hat{O}_2^{(2)}>$ according to the decoupling scheme\ (\ref{decoupling}). Similar
 consideration is given to the boson operators\cite{ng1}. To construct transport equations we consider equations
 of motion for the expectation values $<\Delta_{ij}>, <\chi_{ij}>$ and $<B_{ij}>=<b_ib_j^+>$. The resulting
 (non-linear) equations of motion for
 these functions are linearized and Fourier transformed\cite{ng1}, resulting in a set of linear transport equations
 for the fluctuating variables $\rho_{\vec{k}}(\vec{q},t)=\sum_{\sigma}<c^+_{\vec{k}+\vec{q}/2\sigma}(t)c_{\vec{k}-
 \vec{q}/2\sigma}(t)>$, $\Delta_{\vec{k}}(\vec{q},t)=<c_{-\vec{k}-\vec{q}/2\uparrow}(t)c_{\vec{k}-\vec{q}/2\downarrow}(t)-
 c_{-\vec{k}-\vec{q}/2\downarrow}(t)c_{\vec{k}-\vec{q}/2\uparrow}(t)>$, and $\phi_{\vec{k}}(\vec{q},t)=<b^+_{\vec{k}+
 \vec{q}/2}(t)b_{\vec{k}-\vec{q}/2}(t)>$. An additional equation for the constraint field $\lambda_{\vec{q}}(t)$ is
 obtained by requiring that the density fluctuations of the fermion field is exactly balanced by the density
 fluctuations of the boson field, i.e. $\left(\sum_{\vec{k}}\rho_{\vec{k}}(\vec{q},t)\right)+
 \left(\sum_{\vec{k}}\phi_{\vec{k}}(\vec{q},t)\right)=0$ for all wave-vector $\vec{q}$ and time $t$ in the
 linear-response regime\cite{ng1}.

   Because the equations are linear, the boson variable $\phi_{\vec{k}}(\vec{q},t)$ and constraint field
 $\lambda_{\vec{q}}(t)$ can be eliminated straightforwardly from the coupled equations of motions\cite{ng1}.
 What is perhaps surprising is that the resulting transport equations for $\rho_{\vec{k}}(\vec{q},t)$ and
 $\Delta_{\vec{k}}(\vec{q},t)$ have exactly the form required by Fermi Liquid theory in the $\vec{q}\rightarrow0$
 limit, if we interpret the Fourier transformed variables $\rho_{\vec{k}}(\vec{r},t)$'s and
 $\Delta_{\vec{r}}(\vec{r},t)$ as the local quasi-particle density and Cooper pair amplitudes in Fermi Liquid
 theory, respectively\cite{ng1,leggett}. The exact form of the equation of motion is rather complicated and we shall
 not write it down here. Interested readers can look at Ref.\cite{ng1} for details. The spectrum of the
 quasi-particles is given by the fermion mean-field Hamiltonian\cite{sbtj}
 \begin{equation}
 \label{hmfs}
 H_{MF}^f=\sum_{\vec{k}\sigma}\xi_{\vec{k}}c^+_{\vec{k}\sigma}
 c_{\vec{k}\sigma}+\sum_{\vec{k}}\left[\bar{\Delta}^*(\vec{k})(c_{\vec{k}\uparrow}
 c_{-\vec{k}\downarrow}-c_{\vec{k}\downarrow}c_{-\vec{k}\uparrow})+H.C.\right],
 \end{equation}
 where
 $\xi_{\vec{k}}=-(t\bar{B}+{3J\over8}\bar{\chi})\gamma(\vec{k})+\bar{\lambda}-\mu$, $\bar{B}=<B_{ii+\delta}>$ and
 $\bar{\chi}=<\chi_{ii+\delta}>$ where $\delta=\hat{x},\hat{y}$, and $\gamma(\vec{k})=2(\cos(k_x)+\cos(k_y))$.
 $\bar{\Delta}(\vec{k})={3J\over4}\bar{\Delta}(\cos(k_x)-\cos(k_y))$ represents the quasi-particle pairing field with
 $d_{x^2-y^2}$ symmetry, where $\bar{\Delta}=<\Delta_{i,i+\hat{x}}>$. The mean-field dispersion for the
 quasi-particles is given by $E_f(\vec{k})=\pm\sqrt{\xi_{\vec{k}}^2+|\bar{\Delta}(\vec{k})|^2}$.

  The quasi-particles interact with each other through a temperature, frequency and wave-vector
 -dependent Landau interaction $f_{\vec{k}\vec{p}}(q)$ in the transport equation, with $H_{int}\sim{1\over2}
 \sum_{\vec{k},\vec{p},\vec{q},\omega}\rho_{\vec{k}}(\vec{q},\omega)f_{\vec{k}\vec{p}}(q)
 \rho_{\vec{p}}(-\vec{q},-\omega)$\cite{ng1}, where $q=(\vec{q},\omega;T)$ and
 \begin{equation}
 \label{landaui}
 f_{\vec{k}\vec{p}}(q)=-{3J\over4}\sum_{\mu=\hat{x},\hat{y}}\cos(k_{\mu}-p_{\mu})+V(\vec{q})
 -{1\over\chi_h(q)}(1-2t\chi^c_h(\vec{k};q))(1-2t\chi^c_h(\vec{p};q))+(2t)^2\chi^{cc}_h(\vec{k},\vec{p};q),
 \end{equation}
 where $\chi_h(q)=\sum_{\vec{k}'}\chi_h(\vec{k}',q), \chi_h^c(\vec{k};q)=\sum_{\vec{k}'\mu}\chi_h(\vec{k}',q)
 \cos(k_{\mu}-k_{\mu}')$ and $\chi_h^{cc}(\vec{k},\vec{p};q)=\sum_{\vec{k}'\mu\nu}\chi_h(\vec{k}',q)\cos(k_{\mu}-
 k_{\mu}')\cos(p_{\nu}-k_{\nu}')$, where $\chi_h(\vec{k}',q)={n^b_{\vec{k}'-\vec{q}/2}-n^b_{\vec{k}'+
 \vec{q}/2}\over\omega-\epsilon_{\vec{k}'+\vec{q}/2}+\epsilon_{\vec{k}'-\vec{q}/2}}$.
 $n^b_{\vec{k}}$ are (slave) boson occupation numbers and $\epsilon_{\vec{k}}=-t\bar{\chi}
 \gamma(\vec{k})+\bar{\lambda}$ is the mean-field boson dispersion. $V(\vec{q})$ is the Fourier transform of
 density-density interaction $V(\vec{r})$\cite{ng1}. We find that the Fermi liquid form of transport equation and
 Landau interaction does not change when one goes from the superconducting state to the normal state, and the
 transition is reflected {\em only} in the change in the slave boson response functions $\chi_h,\chi_h^c$ and
 $\chi_h^{cc}$ in Eq.\ (\ref{landaui}).

   To make the expression more tractable we consider the limit $\vec{q}<<\pi/2$ and $k_BT<<t\bar{\chi}$.
 In this limit $\epsilon_{\vec{k}}\rightarrow t\bar{\chi}\vec{k}^2$, $n_{\vec{k}}^b\rightarrow n_{|\vec{k}|}^b,
 \bar{B}\rightarrow x$ and we obtain after some algebra,
 \begin{eqnarray}
 \label{landaui1}
 f_{\vec{k}\vec{p}}(q) & \rightarrow & -{t_{eff}\over\bar{\chi}}\sum_{\mu}(1-z)\cos(k_{\mu})\cos(p_{\mu})
 +V(\vec{q})+2t\sum_{\mu}\left(\cos(k_{\mu})+\cos(p_{\mu})\right)
 \\  \nonumber
 & &
 +{t_{eff}\over\bar{\chi}}\sum_{\mu,\nu}\left({1\over t_{eff}\bar{\chi}}\chi_{h\mu\nu}(q)-\delta_{\mu\nu}\right)
 \sin(k_{\mu})\sin(p_{\nu})
 \end{eqnarray}
 where $t_{eff}=3J\bar{\chi}/4+2tx$ and $z=2tx/t_{eff}$\cite{ng1},
 \[
  \chi_{h\nu\mu}(q)=\left(\chi_{ht}(q)(\delta_{\mu\nu}-{q_{\mu}q_{\nu}\over
  q^2})+\chi_{hL}(q){q_{\mu}q_{\nu}\over q^2}\right)=(2t\bar{\chi})^2\sum_{\vec{k}}k_{\mu}k_{\nu}
  \chi_h(\vec{k},\vec{q})+2t\bar{\chi}x\delta_{\mu\nu}, \]
 is the free (slave) boson current-current response function, $\chi_{hL}(q)$ and $\chi_{ht}(q)$ are the
 corresponding longitudinal and transverse components.  Notice that the bosons contribute only to the
 current-current interaction between quasi-particles in this limit.

  To study the effect of bosons and the superconducting-normal transition we approximate the $\cos(k_{\mu})$
  terms by their average values on the Fermi surface, defined by $\xi_{\vec{k}}=0$\cite{ng1}. The approximation
  keeps the boson contributions exact but simplifies further analysis. We obtain
 \begin{equation}
 \label{landaui2}
 f_{\vec{k}\vec{p}}(q)\rightarrow U_0(q)+{t_{eff}\over\bar{\chi}}\sum_{\mu\nu}\left(F_{1t}(q)
 (\delta_{\mu\nu}-{q_{\mu}q_{\nu}\over q^2})+F_{1L}(q){q_{\mu}q_{\nu}\over q^2}\right)\sin(k_{\mu})\sin(p_{\nu})
 \end{equation}
 where
 \begin{equation}
 \label{flp}
  U_0(q)=V(\vec{q})+aJ+bt, \
  F_{1L(t)}(q)=\chi_{hL(t)}(q)/(t_{eff}\bar{\chi})-1,
  \end{equation}
   where $a, b$ are constants coming from averaging the $\cos(k_{\mu})$ terms on the Fermi surface\cite{ng1}.
  $U_0$ is the usual Landau interaction which couples to density, whereas $F_{1L}$ and $F_{1t}$ couples to
  longitudinal and transverse currents, respectively. Notice that $F_{1L}(q)$ and $F_{1t}(q)$ are identical in
  the limit $\vec{q}\rightarrow0,\omega\neq0$, but are in general different.

    With the simplified form of Landau interaction the density-density and current-current response functions
  can be computed easily. They are given by the standard Fermi-liquid forms\cite{ng1,leggett}
  \begin{equation}
  \label{den}
  \chi_d(\vec{q},\omega)={\chi_{0d}(\vec{q},\omega)\over1-\left(U_0(\vec{q})
  +({F_{1L}(q)\over1+F_{1L}(q)}){\omega^2\over t_{eff}\bar{\chi}\vec{q}^2}\right)\chi_{0d}(\vec{q},\omega)}
  \end{equation}
  and
 \begin{equation}
  \label{current}
  \chi_t(\vec{q},\omega)={\chi_{0t}(\vec{q},\omega)\over1-\left({F_{1t}(q)\over1+F_{1t}(q)}\right)
  {\chi_{0t}(\vec{q},\omega)\over(t_{eff}\bar{\chi})}},
  \end{equation}
  where $\chi_{0d}(q)$ and $\chi_{0t}(q)$ are the mean-field density-density and current-current response functions
  for a BCS superconductor\cite{ng1,leggett} in the absence of Landau interactions, respectively. It is
  straightforward to show that at $T\rightarrow0$ $F_{1t}(q)\rightarrow z-1, F_{1L}(q)\rightarrow
  z\omega^2/(\omega^2-\epsilon_{\vec{q}}^2)-1$ and the
  response functions reduces to the ones obtained in Ref.\cite{ng1}.

  The response functions can be understood in the language of $U(1)$ gauge theory. In the small $x$ limit
  $\chi_{hL}(q)^{-1}$ and $\chi_{ht}(q)^{-1}$ are both of order $x^{-1}$ and we can neglect terms of order O(1)
  compared with them in Eqs.\ (\ref{den}) and \ (\ref{current}). Using Eq.\ ({\ref{flp}) and the identity
  $\chi_{hL}(q)=\omega^2\chi_h(q)/\vec{q}^2$ it is easy to see that
  \[
  \chi_{d(t)}(\vec{q},\omega)\rightarrow{\chi_{0d(t)}(\vec{q},\omega)\chi_{hd(t)}(\vec{q},\omega)\over
  \chi_{0d(t)}(\vec{q},\omega)+\chi_{hd(t)}(\vec{q},\omega)},
  \]
  where the usual Ioffe-Larkin result is recovered\cite{t2,t3}. Recall that in this language, the normal (pseudo-gap)
  state is described by a state with zero bose-condensation, where $\chi_{ht}(0,0)\rightarrow0$ because of
  gauge invariance and the corresponding superfluid density $\rho_s=\chi_t(0,0)$ vanishes\cite{t2,t3}. In the
  corresponding transport equation description, the vanishing of superfluid density is characterized by
  $1+F_{1t}(0,0)\rightarrow0$ (Eq.\ (\ref{flp})). Recall that $1+F_{1t}(q)$ renormalizes (charge) current carried by
  quasi-particles in Fermi liquid theory\cite{leggett} and $1+F_{1t}(0,0)\rightarrow0$ implies that quasi-particle
  carries no current in the long wavelength and zero frequency limit
  $(\vec{q},\omega)\rightarrow(0,0) (t_{eff}\bar{\chi}|\vec{q}|>>\omega)$. The loss of superfluidity in the
  pseudo-gap state is thus not due to renormalization of superconducting gap $\bar{\Delta}\rightarrow0$ in this
  language, but due to vanishing of current carried by quasi-particles. Superconductivity "decouples" from transverse
  electromagnetic field in the DC response of the pseudo-gap phase as a result, and the system becomes a normal metal.

   Our analysis thus suggests that within the framework of Fermi liquid theory, an unconventional phase where (charge)
  current is decoupled from quasi-particles may exist. The phase can be identified naturally as a phase with
  spin-charge separation as suggested in gauge theory approaches\cite{t2,t3,greview,sl} and our analysis provides a
  "Fermi-liquid" definition to this spin-liquid phase. Notice that in the Fermi-liquid description spin-charge
  separation occurs rigorously only in the limit $(\vec{q},\omega)\rightarrow(0,0) (t_{eff}\bar{\chi}|\vec{q}|
  >>\omega)$ and remained coupled at any finite $(\vec{q},\omega)$. In this sense the state we obtained here can be
  characterized as a critical state with "marginal" spin-charge separation. The pseudo-gap phase is  a "marginally"
  spin-charge separated state on top of a BCS-superconductor in this picture but a "marginally" spin-charge
  separated state on top of a normal metal may also exist. In fact it has been suggested that the normal state in
  optimally-doped cuprates is an example of such a state\cite{t2}. The charge dynamics of this state is subtle
  because of its critical nature and will be discussed separately.

  It is also interesting to raise a more general question: within the theoretical framework of Fermi liquid
  superconductor, how many ways can a superconducting to normal transition occur? From Eq.\ (\ref{current})
  (see also Ref.\cite{leggett}), we observe that the superfluid density $\rho_s=\chi_t(\vec{q}\rightarrow0,0)$ may
  vanish either when (i) $\chi_{0t}(\vec{q}\rightarrow0,0)\rightarrow0$, or (ii) when $F_{1t}(0,0)\rightarrow-1$.
  Situation (i) corresponds to the usual BCS mean-field transition or may occur because of strong phase fluctuation
  in the order-parameter wavefunction\cite{t4}. Situation (ii) corresponds to vanishing of current carried by
  quasi-particles and our analysis suggests that this may be what is happening in High-$T_c$ cuprates. There seems
  to be no other possibilities within the framework of Fermi liquid superconductor.

    An advantage of the Fermi liquid description is that it provides a theoretical framework where the pseudo-gap
  phase can be described in terms of only a few Landau parameters, and phenomenologies can be developed without
  referring back to microscopic Hamiltonian(s) where computations are complicated and results are never
  exact anyway. For example, the electromagnetic properties of the pseudo-gap state can be investigated by using a
  phenomenological form of $F_{1t}(q)\sim-1+\chi_d\vec{q}^2+i\omega\sigma(\omega)$. With this form it can be shown
  that the electromagnetic response in the pseudo-gap state is that of a diamagnetic metal with optical
  conductivity $\sigma(\omega)$. Furthermore, strange "vortices" exist in the pseudo-gap\cite{e4} which couples
  only weakly to external magnetic field. We shall discuss consequences of these results separately.

 \acknowledgements
  T.K. Ng acknowledges Prof. P.A. Lee, Prof. X.G. Wen and Dr.Yi Zhou for helpful discussions. This work is supported
  by HKUGC through grant number 602705.

\end{document}